\documentclass{aa}  

\usepackage{graphicx}
\usepackage{txfonts}
\usepackage{lipsum}
\usepackage{subcaption}         
\usepackage{lscape}             
\usepackage{placeins}           

\usepackage{breakcites}
\usepackage{physics}
\usepackage{xcolor}
\usepackage{amsmath}

\providecommand{\der}[2]{\dfrac{{\rm d} #1}{{\rm d} #2}}

\providecommand{\unit}[1]{{\rm #1}}

\usepackage[pdfencoding=auto,breaklinks=true,psdextra]{hyperref}
\hypersetup{
    colorlinks=true,
    linkcolor=blue,
    filecolor=magenta,      
    urlcolor=blue,
    citecolor=blue
}
\usepackage{breakcites}

\begin{document}

\title{Primordial black hole contribution to the stochastic background of gravitational waves}

\author{D. Mart{\'\i}n-Gonz\'alez \thanks{\href{mailto:diegomg@usal.es}{diegomg@usal.es}}}

\institute{Department of Fundamental Physics, University of Salamanca, 37008 Salamanca, Spain}
\abstract{
    {Context.} The amplitude of the detected stochastic gravitational-wave background (SGWB) measured by pulsar timing arrays (PTAs) and the discovery of early and over-massive central black holes at high redshift by the James Webb Space Telescope (JWST) challenge current models of supermassive black hole (SMBH) formation. 
    
    { Aims.} We aim to study if halos containing a significant population of primordial black holes (PBHs) would increase the amplitude of the PTA signal. PBHs add an iso-curvature component to the matter power spectrum, accelerating the formation and merger of dark-matter halos at all redshifts.
    
    { Methods.} We propose that black holes in the halo sink to the center via dynamical friction.  The central black hole grows through hierarchical merging in addition to the gas-accretion channel. 
     We computed the resulting GW amplitude and performed a Bayesian inference analysis using the NANOGrav 15-year dataset.
     
    { Results.} We show that the predicted amplitude of the gravitational-wave background agrees with the observations. Our model only requires $0.09\%-0.12\%$ of the total mass of the halo to fall to the center; this is compatible with a fraction of $f_{\rm pbh}\sim 0.1$ PBHs being dark matter if the in-falling PBHs in the stellar mass range represent about $1\%$ of the total population, as was found in our previous estimation of the formation of SMBHs at $z\sim 6-10$.

    {Conclusions.} The PBH model that explains the JWST new found populations of SMBHs also explain the amplitude of the stochastic background of gravitational waves.}

\maketitle
\nolinenumbers
\section{Introduction}

The detection of the stochastic gravitational-wave background by pulsar timing array (PTA) collaborations has challenged our understanding of the formation of supermassive black holes (SMBHs) \citep{EPTA-IPTA,NanoGrav15yr,PPTA,CPTA}. The measured gravitational-wave amplitude cannot be reconstructed from the local SMBHs' scaling relations unless the local SMBHs are at least ten times more massive than estimated \citep{Sato-Polito2024}. These scaling relations could be extended to higher redshifts thanks to recent discoveries by the James Webb Space Telescope (JWST). The newly found populations host SMBHs with masses in the range of $(10^6-10^8)M_\odot$ at redshifts $6\leq z \leq 10$ with a central BH-to-stellar-mass ratio orders of magnitude greater than the measured local values \citep{Goulding:2023,Bogdan:2024,Maiolino:2024a,Maiolino:2024b,Kovacs:2024,Napolitano2024}. 
\citet{Ellis:2024} proposed that local active galactic nucleus (AGN) populations are a subdominant contribution to the SGWB and that the main contribution comes from SMBHs in local inactive galaxies whose measured scaling relations are more consistent with the JWST high-redshift objects. 

A theoretical explanation for the formation of the newly found populations by the JWST and the deviation from local AGNs' scaling relations is lacking. Several models have been proposed to explain these observations: direct-collapse BHs \citep{Natarajan:2024,Pacucci:2026}, heavy and light seed models \citep{Smith2019,Woods2019,Inayoshi2020,PacucciLoeb:2024,Hu:2025aa}, cosmologically coupled growth \citep{Calza:2024}, and self-interacting dark matter \citep{Shen:2025aa}. Alternatively, these objects could have originated from primordial black holes (PBHs) \citep{Yifu:2024,Yuan:2024,Ziparo:2024aa,Matteri:2025b,Matteri:2025a,Zhang:2025a, Zhang:2025b,DeLuca:2025,Kashlinsky2026, GarciaBellido:2026}. LIGO-Virgo measurements of gravitational-wave emission from merging BHs with low spins and similar masses \citep{Abbott:2016b,GarciaBellido:2021} have brought attention to PBHs as dark matter (DM) candidates \citep{Bird:2016,Kashlinsky:2016,Belotsky:2019}. PBHs are also motivated by the study of  cosmic infrared background anisotropies measured from source-subtracted Spitzer$-$IRAC maps in the near-infrared \citep{Kashlinsky:2018, Hasinger:2020,Cappelluti:2022, Kashlinsky:2025, Kaminsky:2026}.

Different mechanisms have been proposed to account for the formation of PBHs in the early Universe: phase transitions \citep{Hawking1982,Sato:1982,Jedamzik:1997,Khlopov:2010}, modifications of inflation \citep{Garcia-Bellido:1996}, topological defects \citep{Hawking:1989}, among others. Theoretical models predict broad mass functions with peaks in the mass range probed by the LIGO-Virgo observations \citep{Carr:2021}. Scalar-induced gravitational waves would contribute to the PTA-measured amplitude \citep{Afzal:2023}, and this signal would rule out a stellar mass population of  PBHs \citep{Kristiano:2024,Kristiano:2024b,Iovino:2025}. Nevertheless, these estimates are model dependent, and other authors allowed PBHs within this mass range \citep{Riotto:2023,Riotto:2023b,Firouzjahi:2024,Zhu:2024,Gouttenoire:2025,Zhao:2026}. Observationally, microlensing data from the OGLE collaboration have set strong constraints on the PBH abundance \citep{Mroz:2024a,Mroz:2024,Mroz:2025,Mroz:2025b}, but these constraints have been disputed \citep{Hawkins:2025}.

The formation of JWST high-redshift objects was discussed in \citet{Kashlinsky2026}, assuming that a significant fraction of the DM is composed of PBHs in the stellar and substellar mass range. PBHs add an iso-curvature component \citep{Meszaros:1975,Meszaros:1980,Afshordi2003} that favors an earlier collapse of the DM halos and the infall of gas in their potential wells \citep{Kashlinsky2021,Atrio-Barandela2022}. The most massive BHs undergo dynamical friction (DF; \citealt{Chandrasekhar:1943}) and fall to the center. Once they reach the center, DF stops operating. This ``last-parsec problem'' \citep{Begelman:1978,Begelman:1980} is common to all models based on BH mergers, but the clustering of PBHs would help to overcome this limitation \citep{Belotsky:2019,NunoSiles:2025, Stasenko:2025}. Subsequent hierarchical mergers give rise to systems with earlier and more massive central BHs such as those observed with the JWST. We will use this model to compute the amplitude of GW background and compare it with the NANOGrav measurements. 
{
Our main assumptions are that $\left.1\right)$ halos are composed of different fractions of PBHs and dynamical friction drives the most massive to the center of the halos; and $\left.2\right)$ BH growth is driven by hierarchical mergers and gas accretion.}
{
Our approach differs from that of \citet{Ziparo:2024aa}, whose DF timescale does not take into account the effect of the initial angular momentum. Similarly, \citet{Mikage:2025} formulated a model based on PBHs, assuming these seeds grow to become SMBHs, but the authors did not specify a physical model for the BH growth.}
The paper is structured as follows. In Sect.~\ref{sec:GW} we summarize how to compute the amplitude of the gravitational-wave background due to black-hole mergers, in Sect.~\ref{sec:model} we describe our model and calculate the PTA signal, in Sect.~\ref{sec:bayesian} we present and discuss our main results, and in Sect.~\ref{sec:conclusions} we summarize our conclusions.

\section{Gravitational waves from merging BHs.}
\label{sec:GW}
The amplitude of stochastic GWs emitted by merging SMBHs is \citep{Phinney:2001,maggiore2018}
\begin{equation}
h_c^2 (f) = \frac{4 G}{\pi c^2 f^2} \int \frac{\dd z \dd m_1 \dd m_2}{1+z}  \frac{\dd^2 R_{\rm {BH}}}{\dd m_1\dd m_2} \qty(\dv{E_{GW}}{\log f_r})_{f_r=f(1+z)},
\label{eq:hc2}
\end{equation}
where $G$ is the gravitational constant, $c$ is the speed of light, $f$ is the frequency of the gravitational wave, $z$ is the redshift, $m_1$ and $m_2$ are the masses of the merging pair, and $R_{\rm {BH}}$ is the merger rate. The logarithmic derivative of the GW energy for a given chirp mass $\mathcal{M}$ is 
\begin{equation}
      \qty(\dv{E_{GW}}{\log f_r})_{f_r} = \frac{1}{3G}(G\mathcal{M})^{5/3}(\pi f_r)^{2/3}.
\end{equation}
Hereafter, we denote the central BH mass by $m$ and the mass of the host halo by $M$. The SMBH merger rate was computed from the merger rate of halos using the extended Press--Schechter formalism, which generalizes the Press--Schechter formalism using excursion-set technics \citep{Press1974,Bond:1991,Lacey1993}. The halo mass function is 
\begin{align}
      \frac{dn}{dM} = \frac{\rho_0}{M^2}\sqrt{\frac{2}{\pi}}\qty|\dv{\ln \sigma}{\ln M}|\frac{\delta_{c}}{\sigma}\exp\qty{-\frac{\delta_{c}^2}{2\sigma^2}} ,
      \label{eq:PS}
\end{align}
where $\delta_c = 1.686/D(z)$ is the critical value of halo collapse and 
$\sigma(M)$ the variance of the density field at each mass scale, and 
\begin{equation}
D(z) = \frac{5}{2} \Omega_m(z)\left[\Omega^{4/7}_m(z)-\Omega_\Lambda(z)+\left(1+\frac{\Omega_m(z)}{2}\right)\left(1+\frac{\Omega_\Lambda(z)}{70}\right)\right]^{-1}
\end{equation}
is the approximate linear growth factor \citep{Carroll1992}. The probability of merging two halos of masses $M_1, M_2$, with $M_2<M_1$, into a halo of mass $M_f=M_1+M_2$ is
\begin{align}
      \frac{\dd^2 p}{\dd t\dd M_2}
      = \frac{1}{M_f}\sqrt{\frac{2}{\pi}}\qty|\frac{\dot{\delta}_{c}}{{\delta}_{c}}|\qty|\dv{\ln \sigma_f}{\ln M_f}|
      \frac{\delta_{c}}{\sigma_f}\qty(1-\frac{\sigma_f^2}{\sigma_1^2})^{-3/2}
      \nonumber\\
      \times\exp\qty{-\frac{\delta_{c}^2}{2\sigma_f^2 } \qty(1-\frac{\sigma_f^2}{\sigma_1^2})}, 
      \label{eq:extended_PS}
      \end{align}
with $\sigma_f=\sigma(M_f)$ and the dot representing the time derivative.
The merger rate is
\begin{equation}
      \frac{\dd^2 R}{\dd M_1\dd M_2} = \frac{\dd n}{\dd M_1}\frac{\dd^2 p}{\dd z\dd M_2}.
      \label{eq:EPS}
\end{equation}
To relate the SMBHs and the halo-merger rates, we followed \citet{Ellis2023}, assuming that every halo has a central BH and
\begin{equation}
      \frac{\dd^2 R_{\rm BH}}{\dd m_1\dd m_2} \approx \der{M_1}{m_1} \der{M_2}{m_2}\frac{\dd^2 R}{\dd M_1\dd M_2}.
      \label{eq:BHmr}
\end{equation}
We used a semi-analytical model to relate the halo mass and the central BH mass \citep{Barkana2001, Wyithe2003} 
\begin{equation}
      \frac{M}{10^{12} M_\odot} = 10.5 \qty(\frac{\Omega_m(0)}{\Omega_m(z)} \frac{\Delta_c(z)}{18\pi^2})^{-1/2}(1+z)^{-3/2}\qty(\frac{m_{\rm BH}}{10^8 M_\odot})^{3/5},
\label{eq:ConversionLoeb}
\end{equation}
with $\Delta_c(z) = 18\pi^2+82(\Omega_m(z)-1)-39(\Omega_m(z)-1)^2$. Eq.~\eqref{eq:ConversionLoeb} is based on the equilibrium between the radiation absorbed by the gas and the binding energy of the halo. This relation encodes both gas accretion and AGN feedback \citep{PacucciLoeb:2024}.

\section{A model of SMBH formation and growth from PBHs.}
\label{sec:model}
The presence of PBHs favors the formation of structures at small scales. If  $f_{\rm pbh}$ is the fraction of DM as PBHs, the power spectrum is modified with a constant component added to the $\Lambda$CDM spectrum: 
\begin{equation}
      P(k,z)=P_{\rm\scriptsize \Lambda CDM}(k,z)+1.28\times 10^{-6} \left(\frac{D(z)}{D(100)}\right)^2 \qty(\frac{f_{\rm pbh} \overline{m}_{\rm pbh}}{20M_\odot})
      \,\unit{Mpc^3},
\end{equation}
where $\overline{m}_{\rm pbh}=\int \dd m\, m \zeta(m)/\int \dd m \zeta(m)$ and $\zeta(m)$ is the mass function of PBHs. 
\begin{figure}
      \centering
      \includegraphics[width=0.5\textwidth]{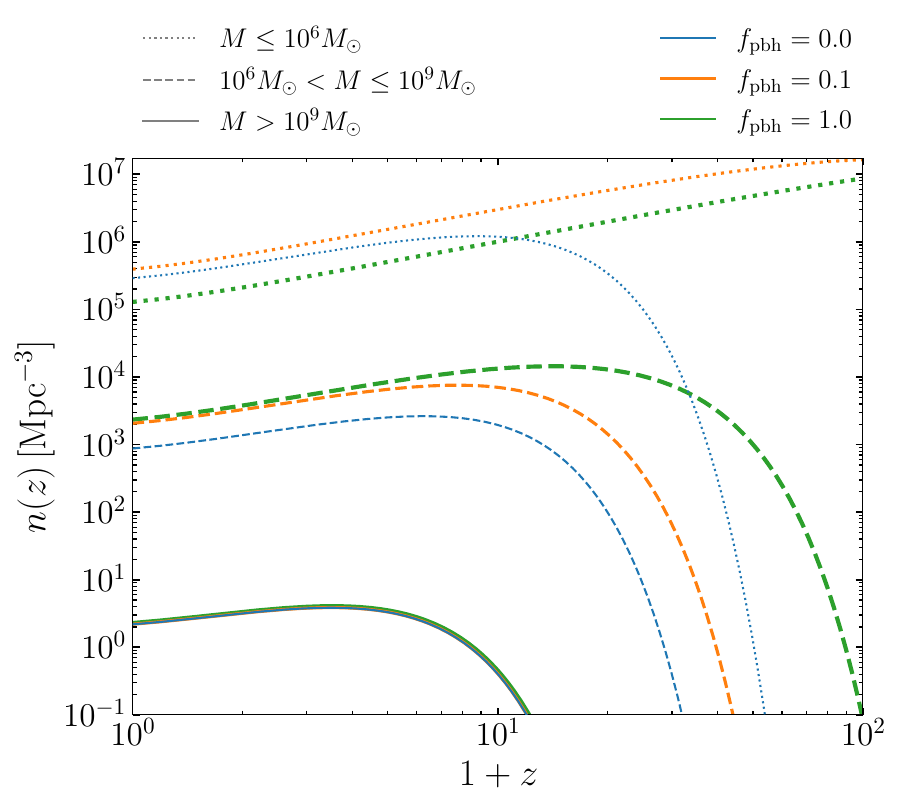}
      \caption{
    Halo number density as a function of redshift. Thin blue, intermediate orange, and thick green solid lines correspond to $f_{\rm pbh} = 0,0.1,1$, respectively. Dotted, dashed, and solid lines correspond to halos in the low-mass, $M\le 10^6$; intermediate, $M=10^6-10^9M_\odot$; and high-mass, $M\ge 10^9$, ranges as indicated.}
      \label{fig:halo-number-density}
\end{figure}
\begin{figure}
      \centering
      \includegraphics[width=0.5\textwidth]{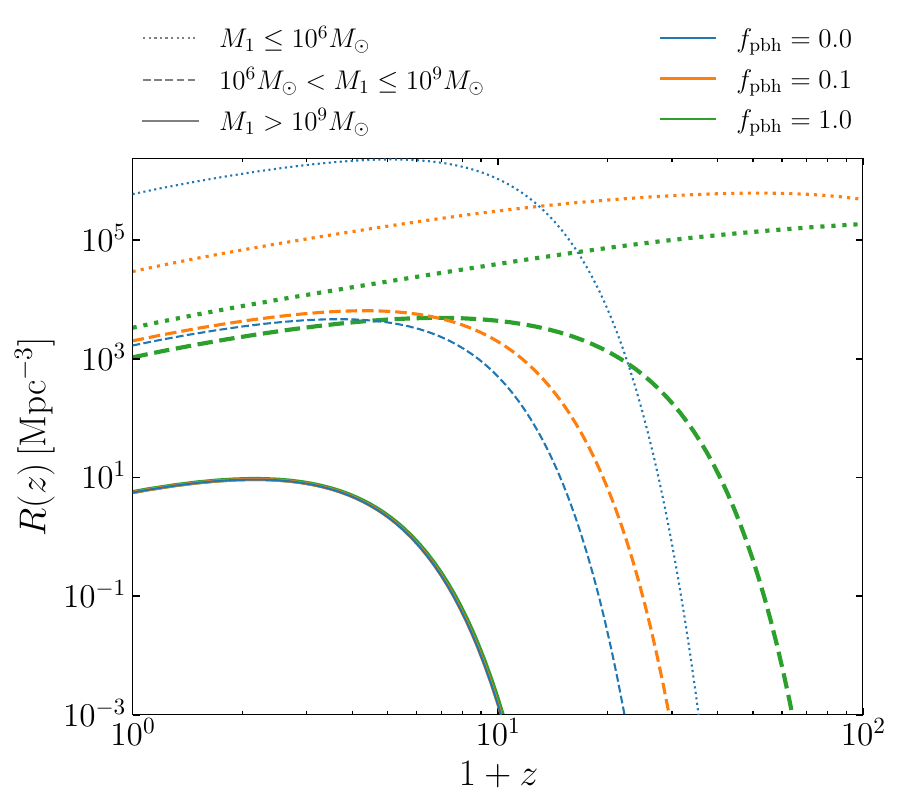}
      \caption{
      Halo-merger rate per unit of redshift. Lines and colors follow the same convention as in Fig.~\ref{fig:halo-number-density}. 
      }
      \label{fig:halo-merger-rate}
\end{figure}

Figure ~\ref{fig:halo-number-density} shows the halo number density, integrated from Eq.~\eqref{eq:PS}, as a function of redshift. Dotted lines correspond to halos with masses of $ M\leq 10^6 M_\odot$, dashed lines to those of $10^6M_\odot\leq M \leq 10^9M_\odot$, and solid lines to those of $M > 10^9M_\odot$. Color and thickness correspond to different fractions: thin blue lines correspond to the standard $\Lambda$CDM case, i.e., $f_{\rm pbh}=0$; orange to $f_{\rm pbh}=0.1$; and the thick green lines to $f_{\rm pbh}=1$. Fig.~\ref{fig:halo-merger-rate} plots the halo-merger rate obtained by integrating Eq.~\eqref{eq:EPS}. Integration of $M_2$ was carried out between $10^{-2}M_1$ and $M_1$. Lines follow the same color convention as in Fig.~\ref{fig:halo-number-density}. Compared to $\Lambda$CDM, the addition of an iso-curvature component accelerates the formation of structures and the merging of halos. In the low- and intermediate-mass ranges, halo collapse could start as early as $z\gtrsim 100$. For $\Lambda$CDM, halo collapse and the subsequent BH mergers would commence much later, between $z\sim 20-30$. For $M > 10^9M_\odot$, the effect of the iso-curvature component is negligible. As we see below, low- and intermediate-mass halos do not contribute significantly to GW production even though their merger rate is the largest. 
The main contribution would come from the halos of $M\geq 10^9M_\odot$, where the merger rate is equal for all the models considered. The difference stems from an earlier BH growth enhanced by dynamical-friction-driven BH mergers, which is only present in the PBH scenario.

In the \citet{Kashlinsky2026} model, 
not all PBHs have sufficient time to reach the center of the halo potential well within a cosmic time, $H^{-1}(z)$. Only BHs with less than $0.1$ of the angular momentum of the virial radius circular orbit ($J\leq 0.1 J_c$) and masses, $m_\bullet$, in the range of $10^{-4}\leq{m}_\bullet \ln\Lambda /M\leq10^{-2}$, where $\ln \Lambda$ is the Coulomb logarithm. In our estimates we fixed it to $\ln \Lambda=10$. 
The growth of the mass of the central BHs, $m_{\rm BH}$, could be split into two contributions:
gas accretion, which we take from Eq.~\eqref{eq:ConversionLoeb}, and BH in-fall due to DF, i.e., ${m_{\rm BH}} = {m_{\rm gas}} + {m_{\rm df}}$. We estimate the evolution of the second term to be
\begin{equation}
      \der{{m}_{\rm df}}{z} = -\frac{H^{-1}(z)}{t_{\rm df}(1+z)} \mathcal{F}(J,{m}_\bullet/M,z) f_{\rm pbh} \dfrac{\Omega_{\rm CDM}}{\Omega_{\rm m}} M,
      \label{eq:mdf}
\end{equation}
where $t_{\rm df}$ is the DF timescale and $\mathcal{F}(J,{m}_\bullet/M,z)$ is the fraction of the most massive BHs with low enough angular momentum that they fall to the center within a cosmic time. 
In \citet{Kashlinsky2026} the fraction of in-falling BHs in halos in our intermediate-mass range, which covers little red dots and UHZ1 type systems, was found to be $\mathcal{F}(J\leq 0.1 J_c, 10^{-5}\leq{m}_\bullet/M\leq 10^{-3})\approx 0.01$. The term $f_{\rm pbh} \frac{\Omega_{\rm CDM}}{\Omega_m} M$ is the total mass of the halo in the form of PBHs. 
The DF timescale is 
\begin{eqnarray}
       t_{\rm df} \approx H^{-1}(z) \sqrt{\dfrac{200}{\Delta_c(z)}}\left(\dfrac{J/J_c}{0.1}\right) \left(\dfrac{0.01}{m_\bullet \ln\Lambda /M}\right).
       \label{eq:timescale}
\end{eqnarray}
This equation incorporates the effect of the angular momentum $J$ \citep{Kashlinsky:1984} and was verified numerically for the Navarro-Frenk-White density profile \citep{Navarro:1996}.

In low-mass halos,  PBHs of $m_\bullet\geq1M_\odot$ fall to center in less than one cosmic time. As halos are assembled into larger systems, their more massive central BHs, of mass ${m}_\bullet$, will now undergo dynamical friction in the next stage of merging. For halos with masses of $M\ge 10^9M_\odot$, the merger rate is drastically reduced as shown in Fig.~\ref{fig:halo-merger-rate}. In these lasts steps of clustering, the massive central BHs merge and dominate the emission of GWs. $\mathcal{F}$ no longer represents the fraction of PBHs with low angular momentum that fall to the center, but the contribution to the final SMBH from the central BHs formed in the previous step of merging. This will introduce a delay in the DF timescale that we can estimate using the merger rate integrated from Eq.~\eqref{eq:EPS}
\begin{equation}
    t_{\rm merg}(z,M) \approx \frac{\dd R/\dd M}{\int dz \,\qty(\dd R/\dd M)} H^{-1}(z) (1+z)^{-1},
\end{equation}
where we are integrating the second halo mass between $10^{-2}M$ and $M$, as in Fig.~\ref{fig:halo-merger-rate}.
Since we probed the largest halo-mass range, $M\geq 10^9M_\odot$, and the angular momenta distribution is unknown, we took $\mathcal{F}(J,{m}_\bullet/M,z)\approx \overline{\mathcal{F}}(J,{m}_\bullet/M)$ as a constant to be determined by the data.
Taking into account the merging time, Eq.~\eqref{eq:mdf} becomes
\begin{equation}
      \der{{m}_{\rm df}}{z} = -\frac{H^{-1}(1+z)^{-1}}{t_{\rm df}+t_{\rm merg}} \mathcal{\overline{F}} f_{\rm pbh} \dfrac{\Omega_{\rm CDM}}{\Omega_{\rm m}} M;
\end{equation}
we can integrate this equation from an effective redshift, $z_{\rm df}$, at which DF starts to operate for the halos. It was treated as a nuisance parameter. At each redshift, $z$, the total mass that fell to the center is now given by
\begin{align}
      m_{\rm BH} = m_{\rm gas} + \overline{\mathcal{F}}  \frac{f_{\rm pbh} \Omega_{\rm CDM}}{\Omega_{\rm m}} M 
      \int^{z_{\rm df}}_{z} \dd\ln(1+z)\frac{H^{-1}}{t_{\rm df} +t_{\rm merg}};
      \label{eq:ourmodel}
\end{align}
where $m_{\rm gas}$ is the gas contribution that can be obtained from Eq.~\eqref{eq:ConversionLoeb}.

\begin{figure}
      \centering
      \includegraphics[width=0.5\textwidth]{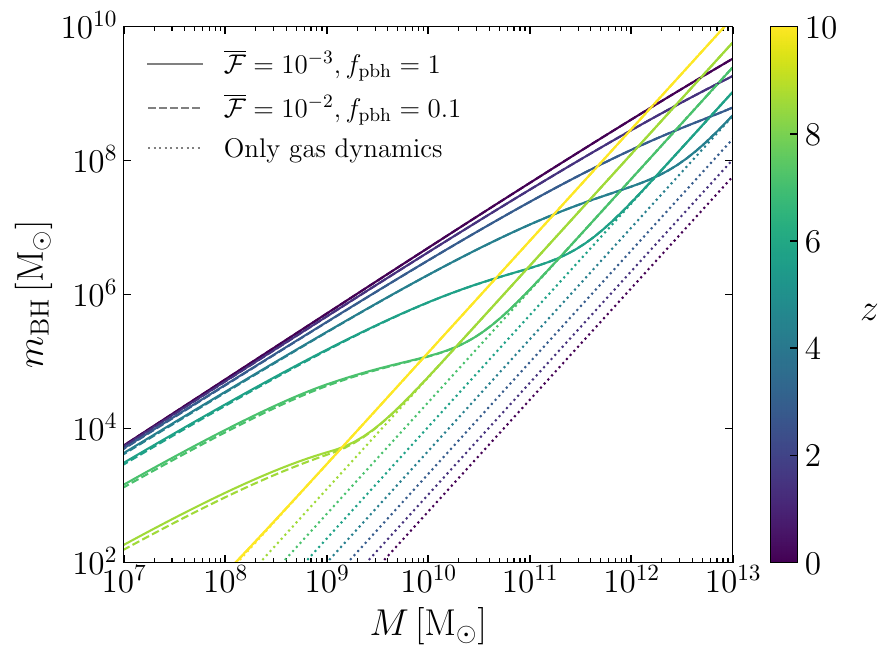}
      \caption{Mass of the central BH versus host halo mass. Solid lines correspond to our fiducial model with $\overline{\mathcal{F}}=10^{-3}$ and $f_{\rm pbh}=1$, dashed lines to $\overline{\mathcal{F}}=10^{-2}$ and $f_{\rm pbh}=0.1$, and both for $z_{\rm df}=20$. Dotted lines correspond to the $\Lambda$CDM model where BHs grow only by accretion, as given in Eq.~\eqref{eq:ConversionLoeb}. The redshift range is $0\leq z\leq 10$ as indicated by the color bar. 
      } 
      \label{fig:mass-conversion}
\end{figure}

\begin{figure}
      \centering
      \includegraphics[width=0.5\textwidth]{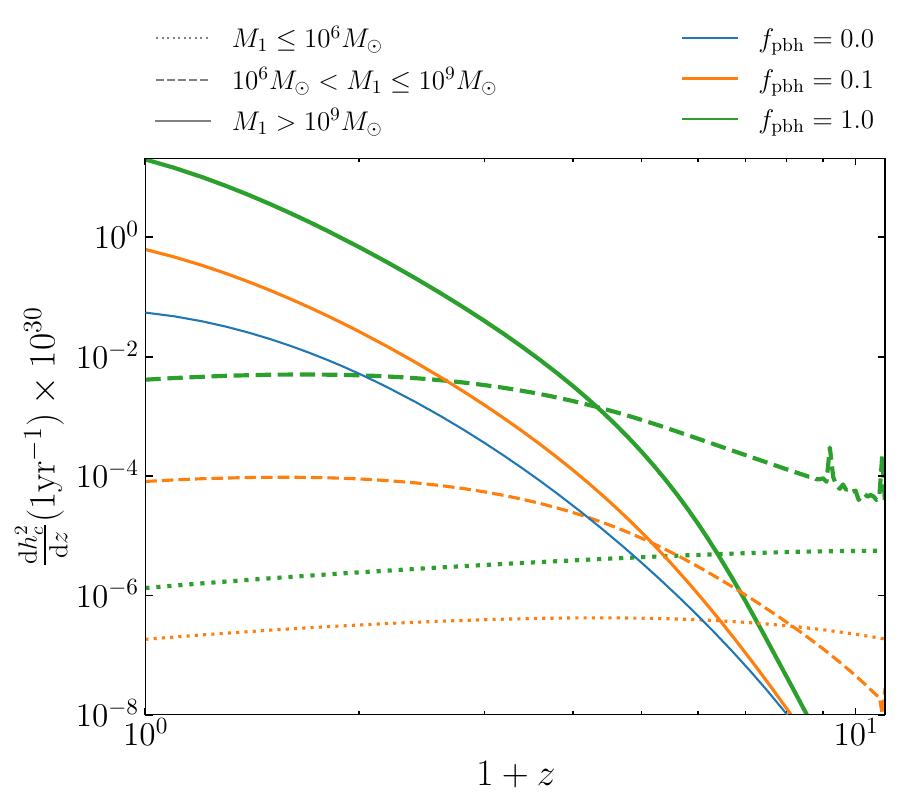}
      \caption{Square of GW amplitude per unit of redshift at a fixed frequency of $1\unit{yr^{-1}}$ as a function of redshift. Lines follow the same conventions as in Fig.~\ref{fig:halo-number-density}. Model parameters were fixed at $\overline{\mathcal{F}}=10^{-3}$ and $z_{\rm df}=20$. }
      \label{fig:dh2dz}
\end{figure}

Figure~\ref{fig:mass-conversion} shows the BH mass as a function of halo mass at different redshifts. Our PBH model is shown for $\overline{\mathcal{F}}=10^{-3}$ and $f_{\rm pbh}=1$ (solid) and $\overline{\mathcal{F}}=10^{-2}$ and $f_{\rm pbh}=0.1$ (dashed), integrated from $z_{\rm df} = 20$. Dotted lines correspond to a model of BH growth with only gas accretion (Eq.~\eqref{eq:ConversionLoeb}). The figure shows that at all redshifts the central BH are more massive in the PBH model than in the $\Lambda$CDM gas-accretion-only model. It also shows that the BH growth due to DF does not depend on $\mathcal{\overline{F}}$ and $f_{\rm pbh}$ separately, but on their product. Fig.~\ref{fig:dh2dz} gives the contribution to the GW amplitude as a function of $z$ for the three halo mass ranges, following the same color convention as in Fig.~\ref{fig:halo-number-density} and integration limits as in Fig.~\ref{fig:halo-merger-rate}, with $\overline{\mathcal{F}}=10^{-3}$, $z_{\rm df}=20$, and $f_{\rm pbh}=0,0.1,1$. Notice that the contribution to the GW amplitude is dominated by the most massive halos, i.e., $M\ge 10^9M_\odot$ at $z\leq5$. If $f_{\rm pbh}=1$, halos in the intermediate mass range contribute $\sim 0.1\%$ to the total GW amplitude, the contribution being much smaller for other PBH fractions or lower mass halos. 
{
Within halos of $M\ge 10^9M_\odot$, the PBH model produces a larger amplitude of GWs than in the $\Lambda$CDM one at all redshifts, as expected from Fig.~\ref{fig:mass-conversion} and as permitted by the hierarchical merger process. As mentioned above, for each halo mass the central BHs in our model are more massive than the central BHs in halos of the same mass of the standard model. Since the amplitude is dominated by the low-redshift SMBHs mergers, the dependence on the choice of $z_{\rm df}$ is negligible.}

\section{Results.}
\label{sec:bayesian}
 
\begin{figure}
      \centering
      \includegraphics[width=0.5\textwidth]{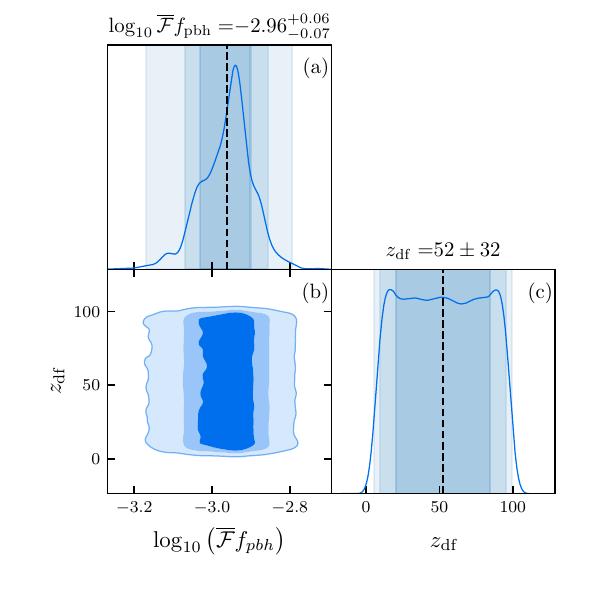}
      \caption{Posterior distributions for the parameters of our fiducial model. The blue regions correspond to the $68\%$, $90\%$, and $99\%$ confidence intervals, respectively. In (a) we show the marginalized posterior distribution for $\log_{10} \overline{F}f_{\rm pbh}$, in (b) the 2D distribution, and in (c) the marginalized posterior distribution for $z_{\rm df}$. The central values of each parameter and their $1\sigma$ errors are given on title and are shown with a dashed black line. }
      \label{fig:posteriors}
\end{figure}

We performed a Bayesian inference analysis to derive the model parameters that best fit the data using the {PTArcade}\footnote{\url{https://andrea-mitridate.github.io/PTArcade/}} code \citep{PTArcade, lamb:2023}. The code fits the free spectrum of the signal,
\begin{equation}
    \rho = \sqrt{\frac{H_0^2 \Omega_{GW}(f)}{8\pi^3 f^5 T_{s}}},\qquad
     \Omega_{\rm GW} (f) = \frac{2\pi^2}{3 H_0^2}f^2 h_c^2(f),
    \label{eq:spectrum}
\end{equation}
to NANOGrav $15\unit{yr}$ of observations \citep{NanoGrav15yr,nanograv:2023}; $\Omega_{\rm GW}(f)$ is the energy density of the GW background, $H_0$ is the Hubble constant, and the period $T_{s}= 16.04\,\unit{yr}$ is the inverse of the minimum data frequency. 
For halos with $M\geq 10^9M_\odot$, $\rho$ depends mainly on the product $\overline{\mathcal{F}} f_{\rm pbh}$ and very weakly on each factor individually, as discussed before. For our statistical analysis  we computed the GW amplitude, $h_c$, by integrating Eq.~\eqref{eq:hc2} from $z=5$ to $z=0$, and we took uniform priors in $\log_{10} \overline{\mathcal{F}} f_{\rm pbh}$ and $z_{\rm df}$, the nuisance parameter of Eq.~\eqref{eq:ourmodel}. The prior intervals were $\log_{10} \overline{\mathcal{F}} f_{\rm pbh}\in [-6,0]$ and $z_{\rm df}\in[5,100]$.
We ran eight chains of $2\times 10^{4}$ samples each. The posterior probability densities are shown in Fig.~\ref{fig:posteriors}. We discarded the first 20\% of samples of our chains to eliminate the bias of the initial values of the parameter estimates
and applied a smoothing scale of $0.2$ using the {GetDist}\footnote{\url{https://getdist.readthedocs.io/en/latest/intro.html}} package \citep{Lewis:2019xzd}.  The blue areas denote the $68\%$, $90\%$, and $99\%$ confidence levels (c.l.s), respectively. Panel (a) shows the marginalized probability density for $\log_{10} \overline{\mathcal{F}}f_{\rm pbh}$ with a mean value of $-2.96^{+0.06}_{-0.07}$ (black dashed line). 
Panel (b) shows the 2D probability density for our parameters. Finally, panel (c) shows the marginalized probability density of $z_{\rm df}$ with a central value of $52\pm 32$. This probability density is dominated by our prior, reflecting that model predictions are rather insensitive to this parameter, as expected.
Since this parameter is difficult to constrain observationally, the GW amplitude being rather insensitive to it shows that our model predictions are robust. 

\begin{figure}
      \centering
      \includegraphics[width=0.5\textwidth]{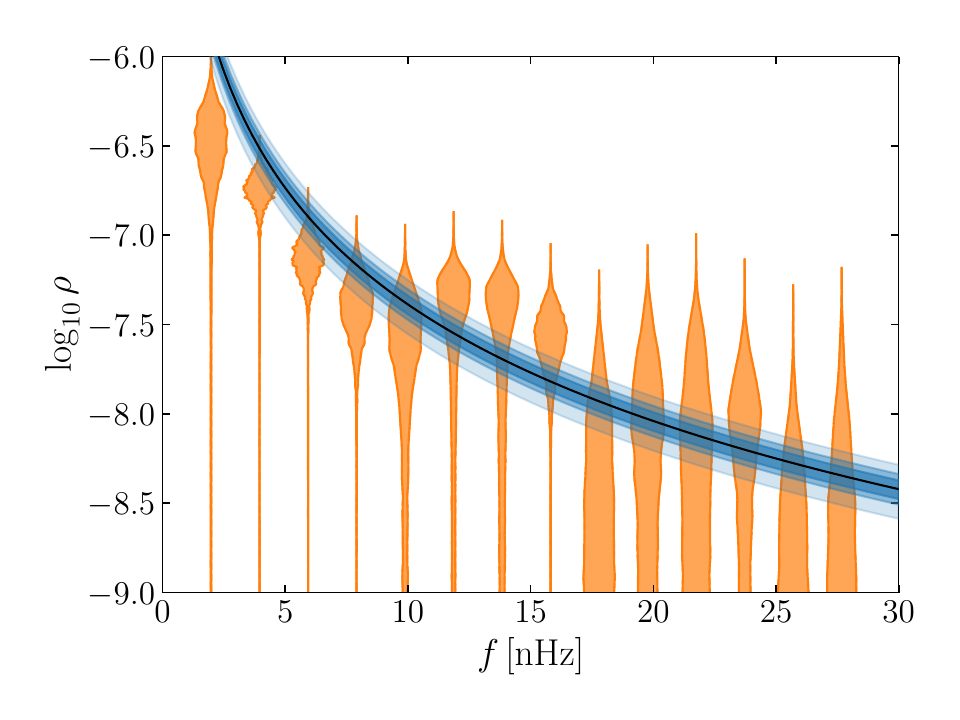}
      \caption{Free spectrum of the GW background versus frequency. The orange shaded areas show the kernel-density estimator for the CURN free spectrum. The best fit to the fiducial model is given by a black line, while the blue areas represent the model confidence intervals. Darker to lighter colors correspond to the $68\%$, $90\%$, and $99\%$ confidence intervals. }
      \label{fig:EnergyDensity}
\end{figure}

In Fig.~\ref{fig:EnergyDensity} we compare the NANOGrav data with our model predictions for different frequencies. The black line shows the free-spectrum-density best fit. The kernel density estimators for the common uncorrelated red-noise (CURN) free spectrum \citep{nanograv:2023} are shown in orange. The dark to light blue areas correspond to the  $68\%$, $90\%$, and $99\%$ c.l.s. 
Our error bars are certainly underestimated compared with works based on empirical scaling relations. Our formalism does not include scatter in the merger rate of BHs in Eq.~\eqref{eq:BHmr} and the dynamical friction time-scale in Eq.~\eqref{eq:timescale} that would add to the error bars.

\begin{figure}
      \centering
      \includegraphics[width=0.5\textwidth]{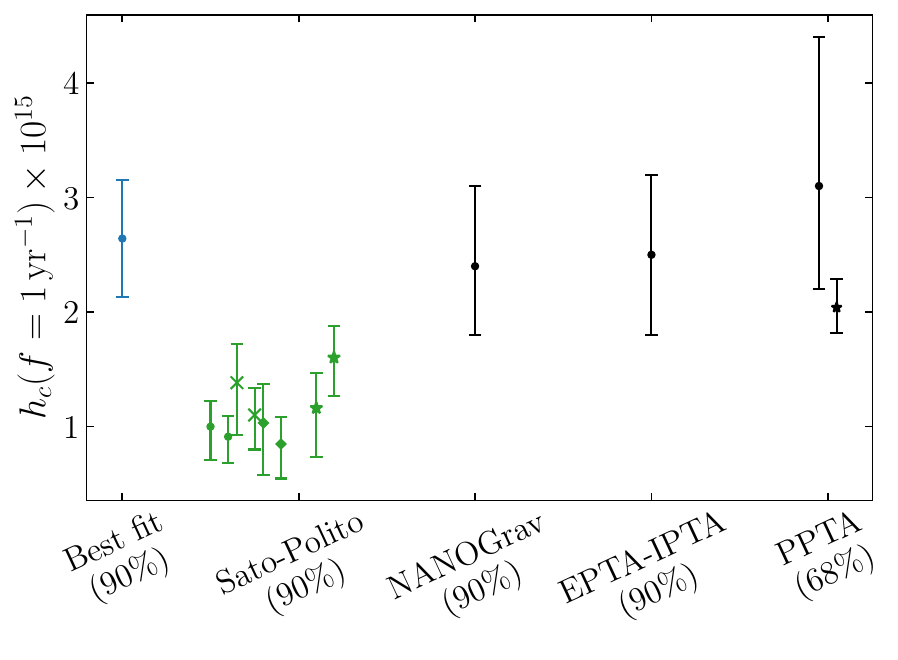}
      \caption{Comparison of GW amplitude for the frequency $1\, {\rm yr}^{-1}$. Our best-fit model and its 90\% c.l. are given in blue. In green we show the theoretical estimated contributions from SMBHs based on different scaling relations \citep{Sato-Polito2024}.  In black, we show the amplitude measurements. NanoGrav and EPTA-IPTA error bars correspond to the $90\%$ c.l. The two PPTA measurements correspond to the $68\%$ c.l.}
      \label{fig:hcComparison}
\end{figure}

Figure~\ref{fig:hcComparison} summarizes our results; it shows the best-fit amplitude of the GW background and its 90\% c.l. predicted by our fiducial PBH model. Other amplitudes correspond to the predictions of SMBH mergers in the concordance model from \citet{Sato-Polito2024}, together with the measured GW amplitude for the $1\, {\rm yr}^{-1}$  NANOGrav, EPTA-IPTA, and PPTA data \citep{EPTA-IPTA,NanoGrav15yr,PPTA}. The observational data are given at the 90\% c.l., except PPTA, which is given at the 68\% c.l. The two values of the PPTA correspond to different assumptions about the frequency dependence of $h_c(f)$. Our model accommodates the amplitude measured by PTAs by requiring only $\sim 0.1\%$ of the halo mass to fall into the center within one cosmic time.


\section{Conclusions.}
\label{sec:conclusions}
Primordial black holes have been proposed as constituents of DM halos. They would add an iso-curvature component to the matter power spectrum at small scales that would accelerate the collapse and growth of halos. Within the halos, they are subject to dynamical friction that drives the most massive BHs to the center, speeding up the formation of the central SMBHs compared to halos of the same mass in the standard cosmological model $\Lambda$CDM. Our model has only physically motivated parameters, but does not require observational scaling relations based on measurements of low-redshift AGNs. 
We calculated the amplitude of the background of GWs in this scenario. 
We show that with a merger-rate history based on the extended Press--Schechter formalism and a semiempirical model of SMBH growth due to PBH mergers and gas accretion, the stochastic background of a GW is dominated by halos of $M_H\ge 10^9M_\odot$ in mass radiating
at $z\leq 5$, similarly to what is seen in the standard model. Nevertheless, the amplitude
of the SGWB is a factor of $\sim 2-3$ larger since BHs in the center
of halos are more massive than in $\Lambda$CDM, as remarked by \citet{Sato-Polito2024}.  The predicted amplitude is independent of the redshift at which the halos start to collapse, $z_{df}$, and is solely determined by the fraction of the most massive PBHs with angular momentum low enough to fall to the center in a cosmic
time at each step in the hierarchical merger process. The fraction of PBHs falling to the center of the halo is the same fraction required to form massive central black holes at $z\approx 10$ if the total PBHs population represents $10\%$ of the DM, as discussed in \citet{Kashlinsky:2025}. Thus, PBHs can explain both the formation of the latest objects found by the JWST and the amplitude of PTAs.

\begin{acknowledgements}
I thank F. Atrio-Barandela, I. De Martino and D. Figueruelo for comments. The PhD grant is supported by
the Fondo Social Europeo Plus, Programa Operativo de Castilla y Le\'on and the Consejer{\'\i}a de Educaci\'on, Junta de Castilla y Le\'on. I also acknowledge financial support from PID2024-158938NB-I00 funded by MICIU AEI/10.13039/501100011033 and by “ERDF A way of making Europe”.
\end{acknowledgements}

\bibliographystyle{aa}
\bibliography{biblio}

\end{document}